\PassOptionsToPackage{unicode}{hyperref}
\PassOptionsToPackage{hyphens}{url}
\PassOptionsToPackage{dvipsnames,svgnames,x11names}{xcolor}
\documentclass[
]{article}
\usepackage{xcolor}
\usepackage{amsmath,amssymb}
\usepackage{iftex}
\ifPDFTeX
  \usepackage[T1]{fontenc}
  \usepackage[utf8]{inputenc}
  \usepackage{textcomp} 
\else 
  \usepackage{unicode-math} 
  \defaultfontfeatures{Scale=MatchLowercase}
  \defaultfontfeatures[\rmfamily]{Ligatures=TeX,Scale=1}
\fi
\usepackage{lmodern}
\ifPDFTeX\else
\fi
\IfFileExists{upquote.sty}{\usepackage{upquote}}{}
\IfFileExists{microtype.sty}{
  \usepackage[]{microtype}
  \UseMicrotypeSet[protrusion]{basicmath} 
}{}
\makeatletter
\@ifundefined{KOMAClassName}{
  \IfFileExists{parskip.sty}{%
    \usepackage{parskip}
  }{
    \setlength{\parindent}{0pt}
    \setlength{\parskip}{6pt plus 2pt minus 1pt}}
}{
  \KOMAoptions{parskip=half}}
\makeatother
\makeatletter
\ifx\paragraph\undefined\else
  \let\oldparagraph\paragraph
  \renewcommand{\paragraph}{
    \@ifstar
      \xxxParagraphStar
      \xxxParagraphNoStar
  }
  \newcommand{\xxxParagraphStar}[1]{\oldparagraph*{#1}\mbox{}}
  \newcommand{\xxxParagraphNoStar}[1]{\oldparagraph{#1}\mbox{}}
\fi
\ifx\subparagraph\undefined\else
  \let\oldsubparagraph\subparagraph
  \renewcommand{\subparagraph}{
    \@ifstar
      \xxxSubParagraphStar
      \xxxSubParagraphNoStar
  }
  \newcommand{\xxxSubParagraphStar}[1]{\oldsubparagraph*{#1}\mbox{}}
  \newcommand{\xxxSubParagraphNoStar}[1]{\oldsubparagraph{#1}\mbox{}}
\fi
\makeatother

\usepackage{color}
\usepackage{fancyvrb}

\DefineVerbatimEnvironment{Highlighting}{Verbatim}{commandchars=\\\{\}}
\usepackage{framed}
\definecolor{shadecolor}{RGB}{241,243,245}
\newenvironment{Shaded}{\begin{snugshade}}{\end{snugshade}}

\newcommand{\AttributeTok}[1]{\textcolor[rgb]{0.40,0.45,0.13}{#1}}

\newcommand{\ConstantTok}[1]{\textcolor[rgb]{0.56,0.35,0.01}{#1}}

\newcommand{\DecValTok}[1]{\textcolor[rgb]{0.68,0.00,0.00}{#1}}

\newcommand{\FunctionTok}[1]{\textcolor[rgb]{0.28,0.35,0.67}{#1}}

\newcommand{\NormalTok}[1]{\textcolor[rgb]{0.00,0.23,0.31}{#1}}

\newcommand{\OtherTok}[1]{\textcolor[rgb]{0.00,0.23,0.31}{#1}}

\newcommand{\SpecialCharTok}[1]{\textcolor[rgb]{0.37,0.37,0.37}{#1}}

\newcommand{\StringTok}[1]{\textcolor[rgb]{0.13,0.47,0.30}{#1}}

\usepackage{longtable,booktabs,array}
\usepackage{calc} 
\usepackage{etoolbox}
\makeatletter
\patchcmd\longtable{\par}{\if@noskipsec\mbox{}\fi\par}{}{}
\makeatother
\IfFileExists{footnotehyper.sty}{\usepackage{footnotehyper}}{\usepackage{footnote}}
\makesavenoteenv{longtable}
\usepackage{graphicx}
\makeatletter
\newsavebox\pandoc@box
\newcommand*\pandocbounded[1]{
  \sbox\pandoc@box{#1}%
  \Gscale@div\@tempa{\textheight}{\dimexpr\ht\pandoc@box+\dp\pandoc@box\relax}%
  \Gscale@div\@tempb{\linewidth}{\wd\pandoc@box}%
  \ifdim\@tempb\p@<\@tempa\p@\let\@tempa\@tempb\fi
  \ifdim\@tempa\p@<\p@\scalebox{\@tempa}{\usebox\pandoc@box}%
  \else\usebox{\pandoc@box}%
  \fi%
}
\def\fps@figure{htbp}
\makeatother

\usepackage[]{natbib}
\usepackage[many]{tcolorbox}
\newtcolorbox[auto counter,number within=section]{codeblock}[2][]{
 fonttitle=\sffamily, lower separated=false,
 attach boxed title to top left={xshift=0.5cm,
        yshift=-2mm},
 enhanced,
 title={Code \thetcbcounter~#2}, breakable,
 label=#1
}

\newenvironment{code}[2][]{\begin{codeblock}[#1,colback=white]{#2}}{\end{codeblock}}
\DefineVerbatimEnvironment{Highlighting}{Verbatim}{xleftmargin=0mm,numbersep=4pt,commandchars=\\\{\}}
\usepackage{setspace}
\AtBeginEnvironment{Shaded}{\onehalfspacing}
\usepackage{arxiv}
\usepackage{orcidlink}
\usepackage{amsmath}
\usepackage[T1]{fontenc}
\makeatletter
\@ifpackageloaded{caption}{}{\usepackage{caption}}
\AtBeginDocument{%
\ifdefined\contentsname
  \renewcommand*\contentsname{Table of contents}
\else
  \newcommand\contentsname{Table of contents}
\fi
\ifdefined\listfigurename
  \renewcommand*\listfigurename{List of Figures}
\else
  \newcommand\listfigurename{List of Figures}
\fi
\ifdefined\listtablename
  \renewcommand*\listtablename{List of Tables}
\else
  \newcommand\listtablename{List of Tables}
\fi
\ifdefined\figurename
  \renewcommand*\figurename{Figure}
\else
  \newcommand\figurename{Figure}
\fi
\ifdefined\tablename
  \renewcommand*\tablename{Table}
\else
  \newcommand\tablename{Table}
\fi
}
\@ifpackageloaded{float}{}{\usepackage{float}}
\floatstyle{ruled}
\@ifundefined{c@chapter}{\newfloat{codelisting}{h}{lop}}{\newfloat{codelisting}{h}{lop}[chapter]}
\floatname{codelisting}{Listing}

\makeatother
\makeatletter
\@ifpackageloaded{caption}{}{\usepackage{caption}}
\@ifpackageloaded{subcaption}{}{\usepackage{subcaption}}
\makeatother
\usepackage{bookmark}
\IfFileExists{xurl.sty}{\usepackage{xurl}}{} 
\makeatletter
\@ifundefined{xmpquote}{\newcommand{\xmpquote}[1]{#1}}{}
\makeatother
\hypersetup{
  pdftitle={Reframing Spatial Uncertainty Visualisation with a Layered Grammar of Graphics},
  pdfauthor={Maggie Xueqi Ma; Weihao Li; Quan Vu; Francis Hui; Emi Tanaka},
  pdfkeywords={\xmpquote{data
visualisation}, \xmpquote{ggplot2}, \xmpquote{glyph
map}, \xmpquote{areal data}, \xmpquote{bivariate palette}},
  colorlinks=true,
  linkcolor={blue},
  filecolor={Maroon},
  citecolor={Blue},
  urlcolor={Blue},
  pdfcreator={LaTeX via pandoc}}

\newcommand{\runninghead}{A Preprint }
\renewcommand{\runninghead}{A Preprint }
\title{Reframing Spatial Uncertainty Visualisation with a Layered
Grammar of Graphics}
\def\asep{\\\\\\ } 
\author{\textbf{Maggie Xueqi Ma}\\Research School of Finance, Actuarial
Studies and Statistics\\Australian National
University\\Canberra,\ 2000\\\asep\textbf{Weihao Li}\\Research School of
Finance, Actuarial Studies and Statistics\\Australian National
University\\Canberra,\ 2000\\\asep\textbf{Quan Vu}\\School of
Mathematics and Statistics\\University of
Sydney\\Sydney,\ 2006\\\asep\textbf{Francis Hui}\\Research School of
Finance, Actuarial Studies and Statistics\\Australian National
University\\Canberra,\ 2000\\\asep\textbf{Emi
Tanaka}~\orcidlink{0000-0002-1455-259X}\\Research School of Finance,
Actuarial Studies and Statistics\\Australian National
University\\Canberra,\ 2000\\\href{mailto:emi.tanaka@anu.edu.au}{emi.tanaka@anu.edu.au}}
\date{}
\begin{document}
\maketitle
\begin{abstract}
Communicating both estimated values and their associated uncertainties
presents a particular challenge in spatial visualisation because spatial
coordinates already occupy the primary visual channels, leaving
relatively few channels available for encoding uncertainty alongside the
estimated values. A range of visualisation techniques has been developed
to address this challenge, including bivariate choropleth maps, pixel
maps, and glyph maps. However, existing implementations typically treat
these techniques as dedicated visualisations rather than expressing them
through the composable elements of a grammar of graphics. We reframe
uncertainty visualisation as an extension of the layered grammar of
graphics, allowing uncertainty-specific representations to be composed,
customised, and extended alongside other graphical components. We
demonstrate this perspective by integrating uncertainty visualisation
into the layered architecture of \texttt{ggplot2}, accommodating these
techniques while also motivating new approaches such as the dual map for
selective attention between estimated values and their associated
uncertainties. We implement the framework in the R package
\texttt{ggincerta} and demonstrate its application through three case
studies.
\end{abstract}
{\bfseries \emph Keywords}
\def\sep{\textbullet\ }
data visualisation \sep ggplot2 \sep glyph map \sep areal data \sep
bivariate palette

\section{Introduction}\label{introduction}

Visualisation of spatial data is a fundamental component of spatial
analysis and communication, enabling the exploration, interpretation,
and dissemination of spatial information. However, spatial data are
often subject to various forms of uncertainty arising from measurement
error, sampling, modelling, and other sources \citep{LI2018313}.
Effectively visualising this uncertainty alongside estimated values is
crucial for supporting informed decision-making and fostering a deeper
understanding of spatial phenomena
\citep{MacEachren_1992, kinkeldeyEvaluatingEffectVisually2017}.

\begin{figure}

\centering{

\includegraphics[width=1\linewidth,height=\textheight,keepaspectratio]{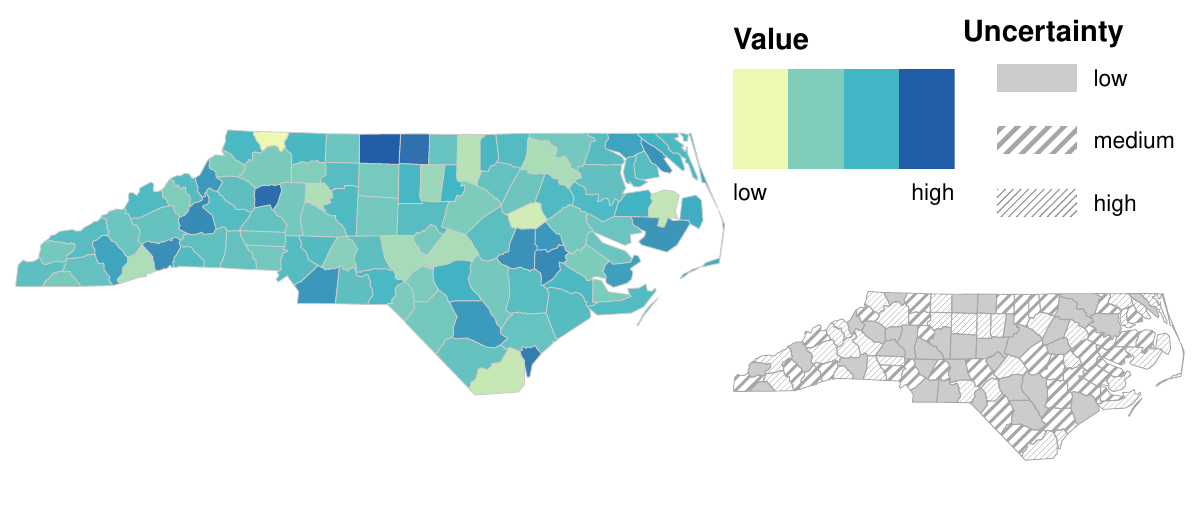}

}

\caption{\label{fig-juxtaposed-map}Juxtaposed visualisation of estimated
values and uncertainty for the North Carolina data. Estimated values are
shown in the main choropleth map, while uncertainty is displayed
separately in the smaller adjacent map.}

\end{figure}

Producing a visualisation involves two related considerations: selecting
an appropriate visualisation method and choosing a tool to implement it.
For the former, one of the most straightforward approaches is to display
uncertainty in a separate plot alongside the estimated values, as shown
in Figure~\ref{fig-juxtaposed-map}. However, such displays may lead
viewers to overlook or underestimate uncertainty and make it difficult
to perceive joint patterns between the estimated value and its
associated uncertainty \citep{correll_value-suppressing_2018}.

We focus primarily on areal data in this paper, although some methods
may also be applicable to other spatial data settings, including
geostatistical (point-referenced) data
\citep{noelaccressieStatisticsSpatialData2015}. In an areal data
setting, space is typically partitioned into a finite set of discrete
units, such as administrative regions, with attribute values measured,
estimated, or reported at the level of these units
\citep{banerjee2014hierarchical}. A central concern is therefore how to
represent and communicate both these regional attribute values and their
associated uncertainty.

Integrating uncertainty directly into the same view provides a natural
alternative, but introduces additional challenges. In spatial maps,
geographic location and areal extent already occupy key visual
attributes commonly used in statistical graphics, including horizontal
and vertical position and area. This leaves fewer options for encoding
additional variables such as uncertainty. Beyond this encoding
constraint, human attention is also limited. When too much information
competes for attention, a display may increase cognitive burden and
become more difficult to interpret \citep{5963660, 6327245}.

Despite these challenges, a substantial body of spatial uncertainty
visualisation methods has been developed, with established taxonomies
describing their diversity \citep{Kinkeldey01112014}. Representative
approaches include bivariate colour schemes, pixelation, glyphs, and
animation. However, this growing collection of methods remains
fragmented, with individual techniques often described in isolation and
without a common language for conceptualising their underlying design
principles. This makes it difficult to identify the visual elements
central to each method and complicates the analysis, comparison, and
interpretation of their respective effects.

At the same time, methods must be translated to tools to produce desired
graphics. A well-designed visualisation tool should not only reproduce a
particular visual output, but also provide a consistent grammatical
structure across different methods, support flexible modification, and
remain compatible and extensible within broader visualisation workflows.
\texttt{ggplot2} is a dominant visualisation tool in R, largely due to
the power of the grammar of graphics that underpins it
\citep{leland_wilkinson_grammar_2005}. This grammatical structure is
applicable across visualisation practices and allows the visual design
of specific methods to be abstracted into more general graphical
objects. Many visualisation packages build on \texttt{ggplot2} to
varying degrees, but not all preserve this grammatical structure. This
can limit users' ability to fully exploit the flexibility and
extensibility of \texttt{ggplot2}, while introducing additional
cognitive and practical burdens.

In this article, we propose the R package \texttt{ggincerta} as a
``native'' approach to implementing spatial uncertainty visualisation
within the \texttt{ggplot2} system. It supports a range of visualisation
methods, including bivariate choropleth maps, pixel maps, and glyph
maps, and introduces a new dual map representation. Its integration with
\texttt{ggplot2} provides a consistent syntax and composable
visualisation components, including layers and scales. This allows all
supported methods to follow the same compositional framework and be
customised using the broader \texttt{ggplot2} ecosystem.

The remainder of the paper is organised as follows.
Section~\ref{sec-related} reviews the development of spatial uncertainty
visualisation methods, their underlying design principles, and their
implementation in existing R packages. Section~\ref{sec-compare}
compares the overall workflow of \texttt{ggincerta} with that of
\texttt{Vizumap} through the construction of a bivariate colour map.
Section~\ref{sec-ggincerta} introduces the visualisation methods
implemented in \texttt{ggincerta}, together with their user-facing
functions, key parameters, and underlying implementation within the
\texttt{ggplot2} system. The methods are demonstrated and discussed
using simulated data that combine different spatial patterns of values
and uncertainty. Section~\ref{sec-application} presents three
representative case studies to demonstrate the broader applicability and
compatibility of the package across different spatial data settings.
Finally, Section~\ref{sec-discuss} discusses the main contributions,
limitations, and directions for future work.

\section{Related work}\label{sec-related}

Visual encoding determines how data attributes are mapped to perceptual
features and can be described in terms of marks and channels: marks
correspond to geometric primitives, while channels represent attributes
through properties such as position, colour, size, and shape
\citep{Bertin2010SemiologyOG, munznerVisualizationAnalysisDesign2014}.
Existing research has therefore explored a range of visual channels for
representing uncertainty, with methods broadly categorised as
colour-based, focus-based, or geometry-based \citep{6327283}.

\begin{figure}

\centering{

\includegraphics[width=1\linewidth,height=\textheight,keepaspectratio]{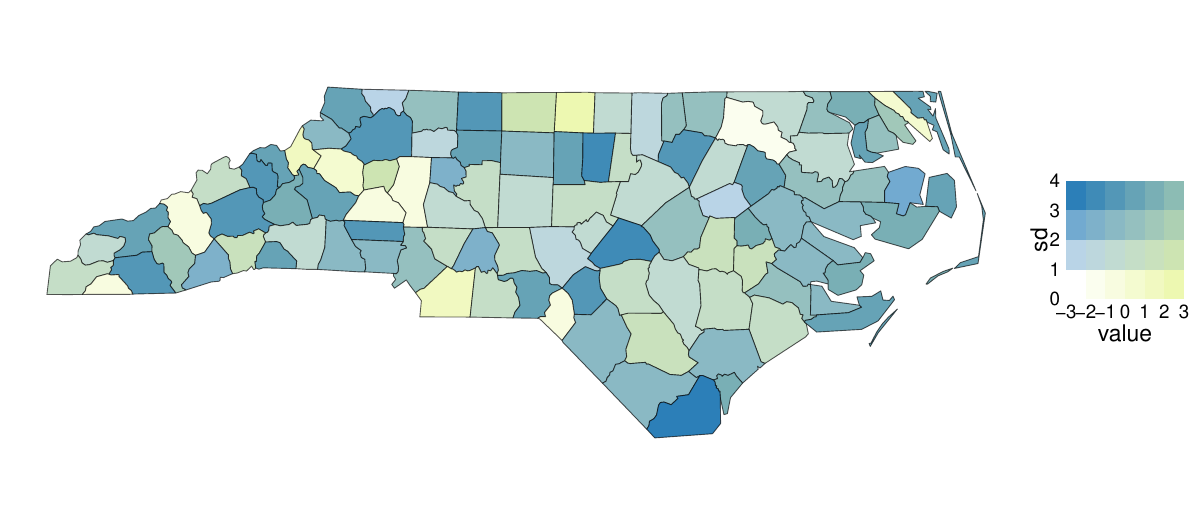}

}

\caption{\label{fig-bivariate-map}Bivariate choropleth map of North
Carolina. The colour grid guide shows the blue-based ramp representing
uncertainty variable (sd), generated independently from a uniform
distribution, and the yellow-based ramp representing the estimated
value, generated independently from a standard normal distribution.}

\end{figure}

Bivariate colour mapping represents two variables jointly within a
single colour palette. For an estimate and its uncertainty, both
typically ordered quantities, a Sequential--Sequential scheme can assign
a separate colour sequence to each variable and combine them to
represent their joint values \citep{Brewer1994, stone_field_2016}.
Figure~\ref{fig-bivariate-map} shows an example based on the North
Carolina shape file from the \texttt{sf} package, using simulated values
and standard deviations.

Several perceptual principles warrant particular consideration when
value and uncertainty are represented together. One is congruence: the
mapping between data values and visual symbols should be consistent with
how those symbols are naturally interpreted. Another is perceptual
separability, which allows each variable to be interpreted with minimal
interference from the other \citep{murrell-hdvw-2026-03-24}. These
considerations are particularly relevant to bivariate colour maps, where
two variables are represented within a shared colour space. For the
bivariate colour scheme used in Figure~\ref{fig-bivariate-map}, a
central challenge is perceptual interference between the two colour
dimensions. When both variables contribute to a blended colour, changes
in one dimension can affect the perception of the other, making their
relative magnitudes difficult to recover independently
\citep{correll_value-suppressing_2018}. This interference can become
more pronounced as the number of classes increases and distinctions
between neighbouring colour combinations become finer. One strategy is
to assign the two variables to visually distinct colour dimensions as
far as possible. For example, hue may be used primarily to encode one
variable, while uncertainty is represented through lightness,
saturation, or opacity. This choice can also improve congruence, as
uncertainty may be represented intuitively by making more uncertain
observations lighter, less saturated, or more transparent
\citep{6327255}. However, complete separability remains difficult
because colour dimensions are perceptually interdependent. Multi-hue
schemes introduce a further trade-off: although hue may help distinguish
the two variables, it is less congruent with ordered data because hue
has no inherent perceptual ordering.

These limitations motivate the use of extrinsic visualisations, which
introduce additional graphical elements such as glyphs rather than
relying only on colour \citep{Kinkeldey01112014}. Since glyphs are
constructed graphical objects, their visual properties can be
deliberately designed to encode uncertainty through channels such as
size, shape, orientation, and geometric distortion \citep{Pang1997}. An
influential early example is the ray-glyph map proposed by
\citet{Carr01011992}, in which the angle of a line segment represents
the estimated value, while a shaded arc attached to the glyph conveys
the associated confidence interval. Subsequent studies have explored a
variety of colour-filled glyph designs and used different visual
attributes to represent uncertainty, including edge crispness
\citep{MacEachren01012005} and rotation angle \citep{Lucchesi2017}.

Many approaches for visualising spatial uncertainty were developed
within the geographic information systems (GIS) and cartographic
literature, where communicating data reliability has long been an
important concern \citep{MacEachren01012005}. The widespread
availability of geographically referenced image data stored on fixed
pixel lattices, such as satellite imagery, also motivated early work on
visualising uncertainty associated with individual pixels
\citep{evansDynamicDisplaySpatial1997a, doi:10.3138/B204-32P4-263L-76W0}.
More broadly, irregular areal units can be divided into regular spatial
tessellations, a common practice in spatial analysis and discrete global
grid systems \citep{Sahr01012003}. Dividing areal units into smaller
spatial elements also creates more opportunities to introduce visual
variation within each region. In a statistical context, values assigned
to these smaller units can be sampled from an underlying distribution,
allowing within-region variation to reflect uncertainty in the original
areal estimates. This idea was proposed and implemented by
\citet{Lucchesi2017} and was subsequently developed into the R package
\texttt{Vizumap}.

\texttt{ggplot2} provides a widely adopted implementation of the layered
grammar of graphics in R \citep{Wickham2010, Wickham2016}, enabling
graphical components to be flexibly combined and extended. Within the R
ecosystem, packages such as \texttt{biscale}
\citep{prener_biscale_2022}, \texttt{multiscales}, \texttt{pixelate}
\citep{taylor_pixelate_2020}, and \texttt{ggdibbler} \citep{Mason2026}
provide implementations of bivariate colour schemes, spatial gridding,
and pixelation-based displays. \texttt{Vizumap} \citep{Lucchesi2021}
additionally provides dedicated plotting functions for several spatial
uncertainty visualisations. However, these packages differ in the extent
to which their functionality is expressed through native
\texttt{ggplot2} components, affecting how individual graphical elements
can be accessed, customised, and composed.

\section{Comparison with Existing Visualisation
Practice}\label{sec-compare}

Before introducing the functions in \texttt{ggincerta}, we use the
construction of Figure~\ref{fig-bivariate-map} as a practical example to
illustrate the effects of grammar-level integration and provide an
initial comparison between different implementation approaches.
\texttt{Vizumap} is a package developed specifically for uncertainty
visualisation of areal data and covers many of the visualisation methods
discussed above, making it a useful representative of tools that use
\texttt{ggplot2} without adopting its formal extension framework.
Although \texttt{Vizumap} already provides a relatively comprehensive
solution for spatial uncertainty visualisation, it uses \texttt{ggplot2}
primarily as a rendering engine. This is a common and effective way
among specialised R visualisation packages, but it adopts a different
workflow from that of a native \texttt{ggplot2} extension and offers
less direct access to the compositional structure of the grammar. The
development of \texttt{ggincerta} was motivated by this difference in
design philosophy: to integrate uncertainty visualisation natively into
the \texttt{ggplot2} ecosystem through a grammar-consistent extension
package.

We can gain an initial impression of the practical benefits of this
shift by comparing the code \ref{code-bivariatemap} used to generate a
bivariate choropleth map in \texttt{ggincerta} with the corresponding
code \ref{code-vizumap} in \texttt{Vizumap}. The most immediate
difference is the amount of code required, particularly the number of
intermediate objects that need to be defined. With \texttt{ggincerta},
users can generate the same result through a single \texttt{ggplot2}
layer expression, whereas \texttt{Vizumap} requires multiple separate
function calls. This simplification also reduces the chance of errors
while improving readability and facilitating debugging.

\begin{code}[code-vizumap]{Bivariate choropleth mapping with Vizumap}

\begin{Shaded}
\begin{Highlighting}[numbers=left,,]
\FunctionTok{library}\NormalTok{(Vizumap)}
\FunctionTok{data}\NormalTok{(nc)}
\NormalTok{nc\_dat }\OtherTok{\textless{}{-}} \FunctionTok{read.uv}\NormalTok{(}\AttributeTok{data =}\NormalTok{ nc, }\AttributeTok{estimate =} \StringTok{"value"}\NormalTok{, }\AttributeTok{error =} \StringTok{"sd"}\NormalTok{)}
\NormalTok{nc\_pal }\OtherTok{\textless{}{-}} \FunctionTok{build\_palette}\NormalTok{(}
  \AttributeTok{name =} \StringTok{"usr"}\NormalTok{,}
  \AttributeTok{colrange =} \FunctionTok{list}\NormalTok{(}
    \AttributeTok{colour =} \FunctionTok{c}\NormalTok{(}\StringTok{"gold"}\NormalTok{, }\StringTok{"red4"}\NormalTok{),}
    \AttributeTok{difC =} \FunctionTok{c}\NormalTok{(}\DecValTok{4}\NormalTok{, }\DecValTok{4}\NormalTok{)))}
\NormalTok{biv\_map }\OtherTok{\textless{}{-}} \FunctionTok{build\_bmap}\NormalTok{(}
  \AttributeTok{data =}\NormalTok{ nc\_dat,}
  \AttributeTok{geoData =}\NormalTok{ nc,}
  \AttributeTok{id =} \StringTok{"NAME"}\NormalTok{,}
  \AttributeTok{palette =}\NormalTok{ nc\_pal,}
  \AttributeTok{terciles =} \ConstantTok{TRUE}\NormalTok{)}
\NormalTok{biv\_key }\OtherTok{\textless{}{-}} \FunctionTok{build\_bkey}\NormalTok{(}\AttributeTok{data =}\NormalTok{ nc\_dat, }\AttributeTok{palette =}\NormalTok{ nc\_pal, }\AttributeTok{terciles =} \ConstantTok{TRUE}\NormalTok{)}
\FunctionTok{attach\_key}\NormalTok{(}\AttributeTok{map =}\NormalTok{ biv\_map, }\AttributeTok{mapkey =}\NormalTok{ biv\_key)}
\end{Highlighting}
\end{Shaded}

\end{code}

Table~\ref{tbl-comparison} divides the overall process into five aspects
to highlight differences in how the two packages construct the
visualisation. Although \texttt{ggincerta} provides functionality
corresponding to the individual steps in \texttt{Vizumap}, these steps
operate as interconnected parts of a unified \texttt{ggplot2} workflow.
Modifications to the data or palette can therefore propagate
consistently throughout the plot. Some steps can also be omitted by
relying on the sensible defaults provided by \texttt{ggplot2}. When
bivariate maps and glyph maps require guide types beyond those provided
by default in \texttt{ggplot2}, \texttt{Vizumap} generates the guides as
separate graphical objects and appends them to the plot using
\texttt{patchwork}. While straightforward, this approach breaks
coherence between the plot and its legend therefore prevents it from
being controlled through the overall theme. By contrast,
\texttt{ggincerta} provides dedicated scale functions with built-in
guides for these map types.

\begin{longtable}[]{@{}
  >{\raggedright\arraybackslash}p{(\linewidth - 4\tabcolsep) * \real{0.3333}}
  >{\raggedright\arraybackslash}p{(\linewidth - 4\tabcolsep) * \real{0.3333}}
  >{\raggedright\arraybackslash}p{(\linewidth - 4\tabcolsep) * \real{0.3333}}@{}}
\caption{Comparison of the workflows of \protect\texttt{Vizumap} and
\protect\texttt{ggincerta}.}\label{tbl-comparison}\tabularnewline
\toprule\noalign{}
\begin{minipage}[b]{\linewidth}\raggedright
Aspect
\end{minipage} & \begin{minipage}[b]{\linewidth}\raggedright
\texttt{Vizumap}
\end{minipage} & \begin{minipage}[b]{\linewidth}\raggedright
\texttt{ggincerta}
\end{minipage} \\
\midrule\noalign{}
\endfirsthead
\toprule\noalign{}
\begin{minipage}[b]{\linewidth}\raggedright
Aspect
\end{minipage} & \begin{minipage}[b]{\linewidth}\raggedright
\texttt{Vizumap}
\end{minipage} & \begin{minipage}[b]{\linewidth}\raggedright
\texttt{ggincerta}
\end{minipage} \\
\midrule\noalign{}
\endhead
\bottomrule\noalign{}
\endlastfoot
\textbf{Data preparation} & Dedicated \texttt{read.uv()} function &
Direct input to \texttt{ggplot()} \\
\textbf{Data manipulation} & Separate \texttt{build\_*()} functions &
Integrated scales and stats \\
\textbf{Scales} & Separate palette and key construction &
\texttt{ggplot2} scale functions \\
\textbf{Composition} & Manual map--key assembly & Integrated layers and
guides \\
\textbf{Workflow} & Sequential function calls & Layered grammar of
graphics \\
\end{longtable}

Beyond individual functions and code, grammatical integration also
clarifies the underlying logic of the visualisation workflow. Users can
understand how a plot is constructed through concepts and skills already
familiar to much of the R visualisation community, rather than treating
an unfamiliar specialised function as a black box.

\section{Methods for spatial uncertainty
visualisation}\label{sec-ggincerta}

We designed the R package \texttt{ggincerta} to map value and
uncertainty to a range of visual representations through native
integration with \texttt{ggplot2}. This allows uncertainty
visualisations to be constructed within standard \texttt{ggplot2}
workflows, retaining functionality such as transformations, faceting,
and theming. At the time of writing, \texttt{ggincerta} supports
bivariate colour, pixel, and glyph-based representations, and introduces
a dual map representation. All methods presented below are static
visualisations. Dynamic approaches like animation fall outside the scope
of \texttt{ggincerta}.

\subsection{Bivariate choropleth map}\label{bivariate-choropleth-map}

Bivariate choropleth maps represent two variables jointly through a
two-dimensional colour scheme. In Figure~\ref{fig-bivariate-map}, the
two variables define the data space, with their combinations determining
the colour encoding. Unlike the conventional mapping of one variable to
one aesthetic in \texttt{ggplot2}, \texttt{ggincerta} uses the
\texttt{duo()} function to bundle the estimated value and its
uncertainty into a single list-based object mapped to \texttt{fill} or
\texttt{colour}. The \texttt{scale\_*\_bivariate} family then
discretises the two variables and assigns a colour to each resulting
category through the selected palette.

Code block \ref{code-bivariatemap} demonstrates the basic workflow for
generating Figure~\ref{fig-bivariate-map} in \texttt{ggincerta}. The
\texttt{ggplot} call and \texttt{geom\_sf} specify the input data and
aesthetic mappings for constructing the choropleth map layer. The class
of this object allows \texttt{ggplot2}'s scale-dispatch mechanism to
select \texttt{scale\_fill\_bivariate()} automatically, without
explicitly specifying the scale.

\begin{code}[code-bivariatemap]{Bivariate choropleth mapping with ggincerta}

\begin{Shaded}
\begin{Highlighting}[numbers=left,,]
\FunctionTok{ggplot}\NormalTok{(nc) }\SpecialCharTok{+}
  \FunctionTok{geom\_sf}\NormalTok{(}\FunctionTok{aes}\NormalTok{(}\AttributeTok{fill =} \FunctionTok{duo}\NormalTok{(value, sd)))}
\end{Highlighting}
\end{Shaded}

\end{code}

As illustrated in Figure~\ref{fig-bivariate-map}, bivariate colour
mapping retains the geometric structure of a conventional choropleth map
and therefore does not require a new \texttt{Geom}. Instead, its
distinctive operation is the mapping of paired variables to a single
colour aesthetic. This is implemented through specialised scales and
guides, while the existing \texttt{geom\_sf()} continues to provide the
geometric layer.

One of the key contributions of \texttt{ggincerta} as a visualisation
tool is its increased flexibility, achieved by expanding the range of
visual configurations exposed through the underlying
\texttt{bivariate\_scale()} constructor and user-facing functions such
as \texttt{scale\_fill\_bivariate()}. Broadly, these controls can be
grouped into two aspects: data transformation, palette and guide
control.

\subsubsection{Data binning and
transformation}\label{data-binning-and-transformation}

Following the discrete combinatorial structure of bivariate mapping,
\texttt{guide\_bivariate()} is represented as a grid with four bins per
variable by default. However, the actual number of bins may vary, as
shown in Figure~\ref{fig-bivariate-map} where the value variable is
divided into six groups. This occurs because \texttt{bivariate\_scale()}
uses \texttt{scales::breaks\_pretty()} to generate interpretable
breakpoints for constructing equally spaced bins. The argument
\texttt{n\_breaks} specifies only the desired number of intervals, so
the actual number may vary depending on the data range. Users can
disable this behaviour with \texttt{nice.breaks\ =\ FALSE}, use
quantile-based breakpoints with \texttt{quantile\ =\ TRUE}, or directly
control the partitioning and data range through the \texttt{breaks} and
\texttt{limits} arguments. In addition to controlling bin boundaries,
users can apply transformations through the \texttt{transform} argument,
with binning performed on the transformed values.

\subsubsection{Palettes and guides in
scales}\label{palettes-and-guides-in-scales}

\begin{figure}

\centering{

\includegraphics[width=1\linewidth,height=\textheight,keepaspectratio]{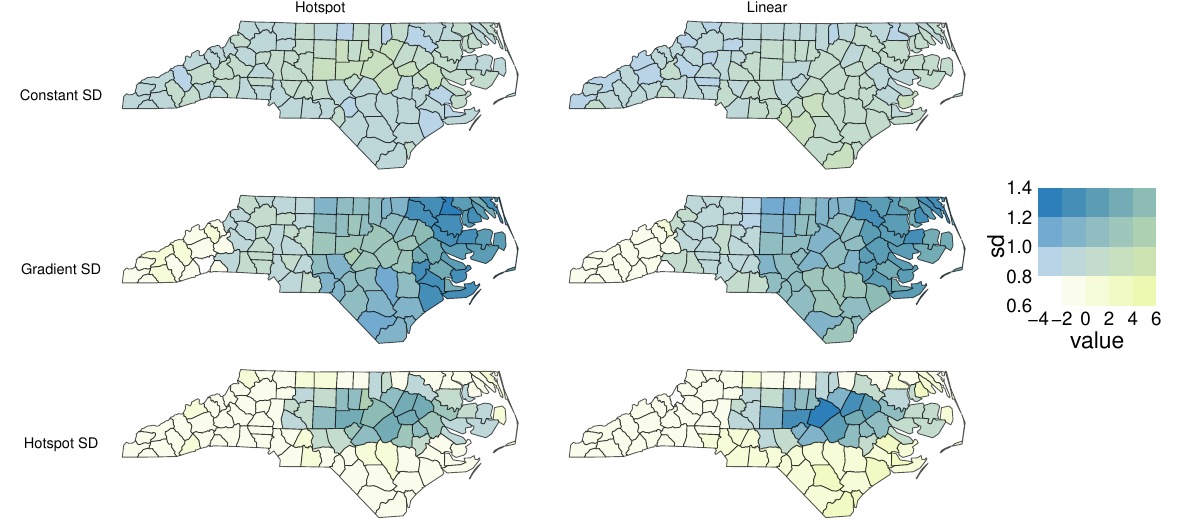}

}

\caption{\label{fig-bimap-sim}Faceted plots showing combinations of
spatial signal and standard deviation patterns. Columns represent
hotspot and linear trend patterns in the underlying spatial signal,
while rows represent the spatial patterns in the standard deviation of
the Gaussian distributions from which county-level values are
simulated.}

\end{figure}

Another important consideration in bivariate palette design is the
balance of visual emphasis between value and uncertainty. Uncertainty
usually serves as supplementary information about the reliability or
confidence associated with the primary data variable. To illustrate this
issue, Figure~\ref{fig-bimap-sim} presents a simulated dataset combining
two spatial signal patterns, a directional trend and a localised
hotspot, with three uncertainty patterns: a constant level, a
west-to-east gradient, and a localised hotspot. The signal trend
increases towards the southeast, while the uncertainty gradient
increases from west to east. The signal and uncertainty hotspots share
the same location. For county \(i\), we simulate a response
\(Y_i = \mu_i + \varepsilon_i\), where
\(\varepsilon_i \sim N(0,\sigma_i^2)\), \(\mu_i\) represents the
underlying signal, and \(\sigma_i\) follows the specified uncertainty
pattern.

When uncertainty is constant (top row of Figure~\ref{fig-bimap-sim}),
the underlying signal may remain perceptible, although its visibility
varies across spatial patterns and may be reduced compared with a
univariate choropleth map. With a fixed standard deviation, the mapped
colours are restricted to a single row of the bivariate colour grid,
narrowing the available colour range and potentially reducing visual
contrast. When uncertainty itself exhibits a spatial pattern (middle and
bottom rows), it introduces additional colour variation that may compete
with the underlying signal, potentially obscuring or confounding its
spatial structure.

Consequently, both the visual channels used to encode uncertainty and
the manner in which they are integrated with value representations
should be adapted to the analytical and interpretive goals of the
visualisation work. This idea is closely related to the ``focusing''
metaphor proposed by \citet{MacEachren_1992}, where uncertain
information is visually pushed ``out of focus'' through reductions in
saturation, clarity, or visual intensity, while more certain information
remains visually prominent.

\begin{figure}

\centering{

\includegraphics[width=1\linewidth,height=\textheight,keepaspectratio]{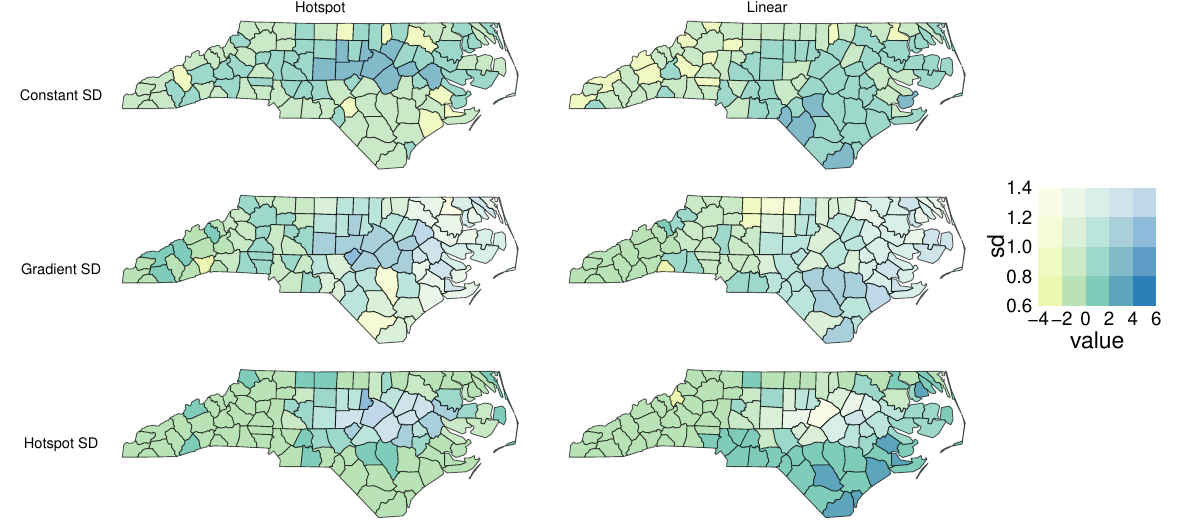}

}

\caption{\label{fig-fade-sim}Faceted bivariate choropleth maps using
colour fade palette on simulated data.}

\end{figure}

Subsequent development in \texttt{ggincerta} expanded this design by
supporting a broader range of bivariate palette methods. These
extensions are implemented through \texttt{bivar\_fade\_palette()}
function, which can be supplied via the \texttt{palette\_fun} argument
in \texttt{bivariate\_scale()}. Rather than directly blending two colour
ramps, \texttt{bivar\_fade\_palette()} first generates a single colour
ramp from the supplied colours as the value aesthetic, and then
attenuates the intensity of a specified visual property along the
uncertainty dimension. By default, uncertainty is encoded through
variations in lightness, although transparency and desaturation can also
be used through the \texttt{fade} argument, with available options
including \texttt{"lighten"}, \texttt{"alpha"}, and
\texttt{"desaturate"}. In addition, \texttt{bivar\_fade\_palette()}
allows colours to be specified independently for each value bin,
providing a more flexible and practical approach to bivariate colour
mapping. Compared with the previous colour blending palette,
Figure~\ref{fig-fade-sim} shows that the colour fading palette provides
a better balance between the perceptual strength of the two variables.
When uncertainty is spatially constant and remains at a moderate level,
sufficient colour contrast is retained so that the underlying signal
pattern is not overly attenuated. Across the simulated combinations, one
empirical observation is that when the spatial patterns of the signal
and uncertainty are less aligned, each tends to remain more visually
distinguishable. In contrast, stronger spatial overlap can obscure the
signal more substantially. This behaviour is consistent with the
intended role of uncertainty visualisation: where uncertainty is high,
the visual prominence of an unstable spatial signal should be reduced so
that its limited reliability is communicated more clearly.

Beyond these two automatically generated palettes within
\texttt{bivariate\_scale()}, \texttt{ggincerta} also provides a manual
palette interface that allows users to fully control the mapping by
supplying the complete set of colours for all bivariate combinations.
The underlying \texttt{manual\_bivariate\_scale()} constructor is
exposed through user-facing functions such as
\texttt{scale\_fill\_bivariate\_manual()}, whose usage is analogous to
the manual scale functions in \texttt{ggplot2}.

\begin{figure}

\centering{

\includegraphics[width=1\linewidth,height=\textheight,keepaspectratio]{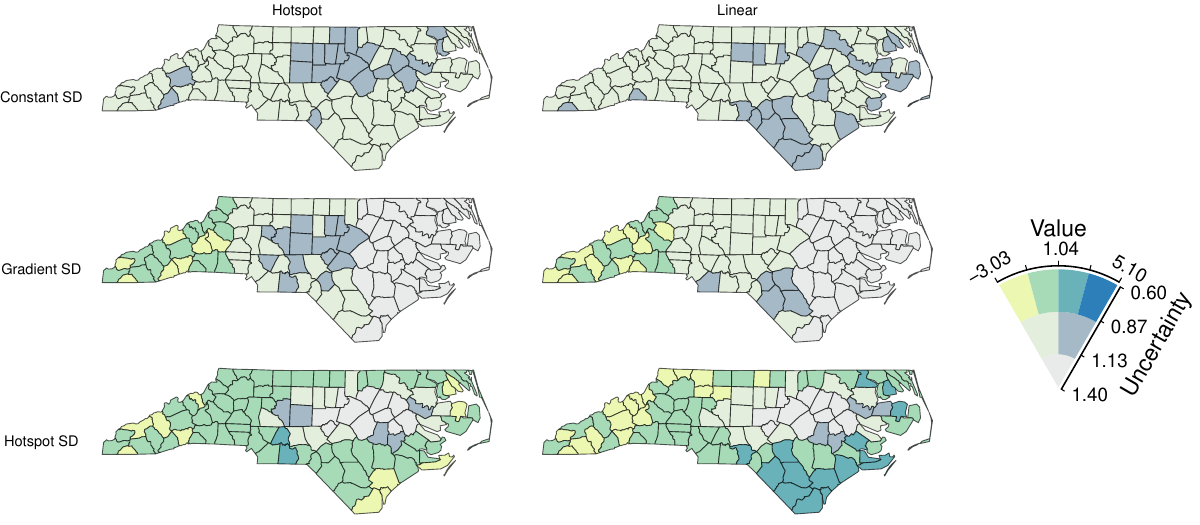}

}

\caption{\label{fig-vsup-map}Faceted bivariate choropleth maps using
VSUP on simulated data.}

\end{figure}

Another approach specifically designed for uncertainty visualisation is
the Value-Suppressing Uncertainty Palette (VSUP), where distinctions
between values are progressively reduced at higher uncertainty levels in
order to direct perceptual attention toward more reliable regions
\citep{correll_value-suppressing_2018}. As shown in
Figure~\ref{fig-vsup-map}, the visual appearance differs from the
classical bivariate colour map in Figure~\ref{fig-bivariate-map} through
the use of a fan-shaped legend that directly reflects the underlying
division of data space. Uncertainty is encoded along the radial
direction, while the granularity of value representation decreases
toward the inner rings and increases toward the outer rings.
Consequently, lower uncertainty regions retain a broader range of
distinguishable colours, whereas higher uncertainty regions gradually
converge toward grey tones. This design encourages viewers to
perceptually deemphasise unreliable information and focus on more
trustworthy regions.

\texttt{ggincerta} provides \texttt{scale\_fill\_vsup()} to implement
this palette method together with its corresponding guide. The
\texttt{layers} argument controls the number of uncertainty levels,
while \texttt{branch} specifies the number of value subdivisions
introduced at each successive layer. The \texttt{colours} argument in
\texttt{scale\_fill\_vsup()} specifies the colours used for the lowest
uncertainty level. If the number of supplied colours differs from the
baseline number of value categories, \texttt{vsup\_palette()}
interpolates the colours to generate the base palette.

Guides are naturally generated alongside scales and serve as
interpretative devices that help viewers establish the relationship
between aesthetic mappings and the underlying data, such as axes for
position scales and legends for colour and fill scales in
\texttt{ggplot2} \citep{Wickham2016}. In \texttt{ggincerta},
\texttt{guide\_bivariate()} and \texttt{guide\_vsup()} work alongside
their corresponding scales to communicate the joint mapping of value and
uncertainty to colour.

\subsection{Pixel map}\label{pixel-map}

In a pixel map, each areal unit is subdivided into smaller cells, with
values sampled from a distribution parameterised by the corresponding
estimate and uncertainty. Hexagonal pixels are used by default in
\texttt{ggincerta} for their isotropic properties and reduced visual
interference \citep{Carr01011992}, with square and rectangular
alternatives also supported. As this process primarily involves spatial
tessellation and grouped random sampling, \texttt{ggincerta} implements
it through a specialised \texttt{StatPixel} within
\texttt{geom\_sf\_pixel()}. The \texttt{distribution} argument supports
normal and uniform distributions, while the \texttt{seed} argument
ensures reproducibility. For colour mapping, \texttt{duo\_pixel()}
bundles value and uncertainty into a single mapping object and
automatically invokes \texttt{scale\_fill\_pixel()}, providing a
continuous colour scale consistent with
\texttt{scale\_fill\_distiller()}.

\begin{figure}

\centering{

\includegraphics[width=1\linewidth,height=\textheight,keepaspectratio]{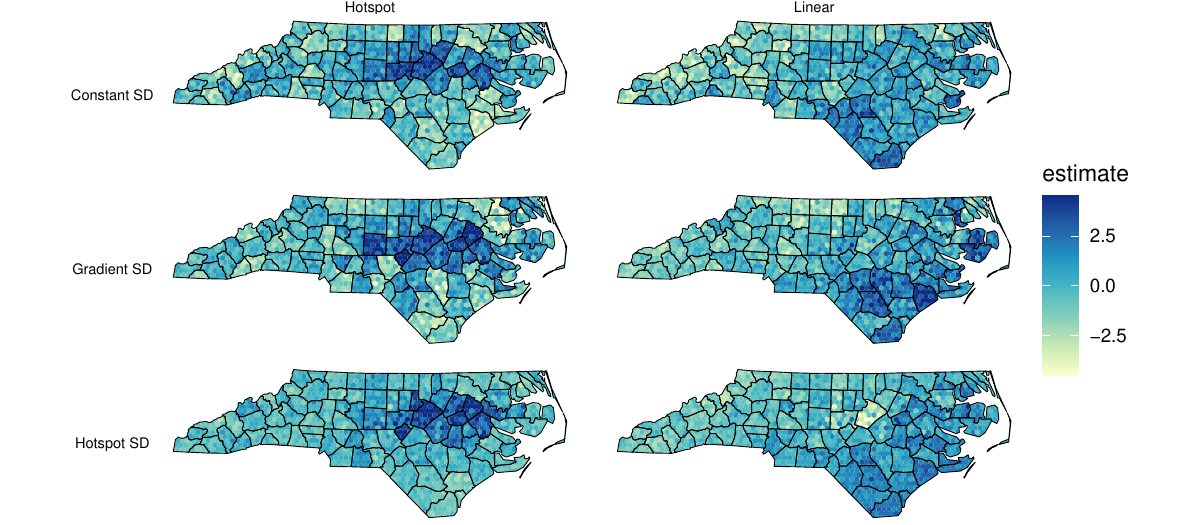}

}

\caption{\label{fig-pixel-map}Faceted pixel maps on simulated data}

\end{figure}

Figure~\ref{fig-pixel-map} produces a markedly different perceptual
balance from the colour-based methods above. The primary variable
largely determines the resulting visualisation, while standard deviation
is encoded more implicitly through perturbations rather than through a
separate visual channel. As a result, uncertainty has a more limited
influence on the overall appearance of the map. When uncertainty is
constant, the perturbation is distributed relatively uniformly across
space, leaving the display visually similar to a conventional choropleth
map. When the signal and uncertainty patterns overlap, the signal is
still attenuated to some extent, although substantially less than in the
preceding bivariate colour maps. The trade-off is that uncertainty
becomes much less directly interpretable. Although local regions of
higher uncertainty can sometimes be identified, the broader spatial
pattern is difficult to assess from the display alone.

\subsection{Glyph map}\label{glyph-map}

\begin{figure}

\centering{

\includegraphics[width=1\linewidth,height=\textheight,keepaspectratio]{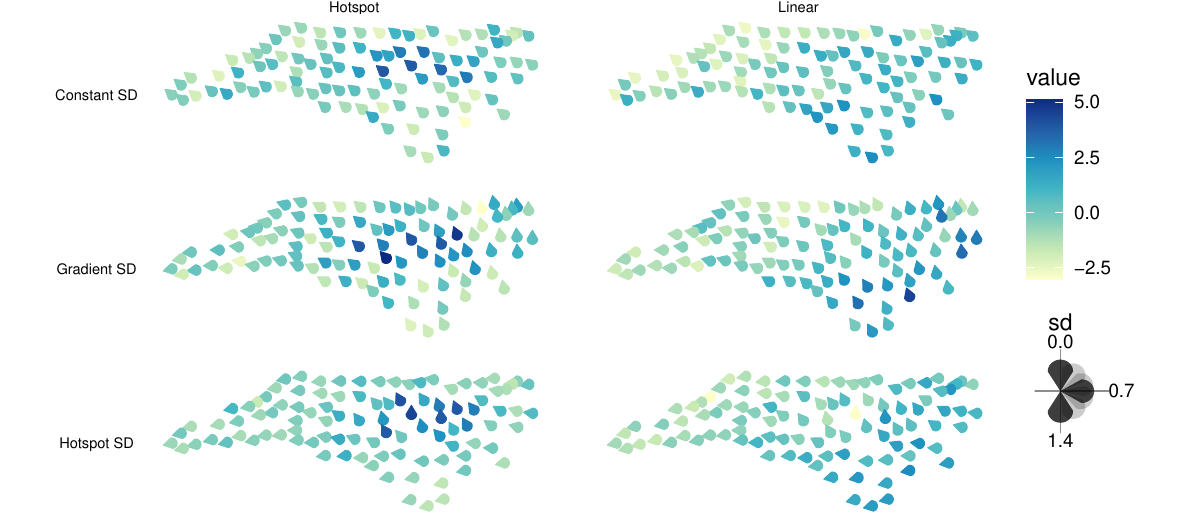}

}

\caption{\label{fig-drop-sim}Faceted glyph maps using drop-shaped glyphs
on simulated data. Colour represents the simulated value, while the
rotation angle of each drop glyph represents the standard deviation.}

\end{figure}

A glyph map (Figure~\ref{fig-drop-sim}) is essentially a point-based map
in which each spatial region is represented by a glyph positioned at a
representative location, such as its centroid. This abstraction can
reduce perceptual bias introduced by irregular administrative boundaries
and unequal region sizes, which was one of the motivations behind
earlier glyph map designs \citep{Lucchesi2017}. In \texttt{ggincerta},
\texttt{geom\_sf\_glyph()} supports multiple glyph designs with
uncertainty encoded through properties such as rotation angle and facial
expression, while remaining compatible with bivariate colour scales and
other uncertainty-aware mapping components.

Unlike pixelation, constructing glyphs involves extending both the
statistical transformation and geometric representation of a layer.
Accordingly, \texttt{ggincerta} defines \texttt{StatSfPin} and
\texttt{GeomSfPin}: the former generates glyph geometries from spatial
objects and aesthetic mappings, while the latter integrates these
geometries into the \texttt{ggplot2} rendering pipeline.

\subsubsection{Glyph designs}\label{glyph-designs}

Regular-shaped glyphs, including circles, squares, triangles, and
hexagons, are provided as additional representations in
\texttt{ggincerta}. By default, \texttt{geom\_sf\_glyph()} converts each
areal geometry into a representative point using the user-specifiable
\texttt{point\_fun} argument. The resulting layer functions similarly to
\texttt{geom\_point()} and supports bivariate colour mapping through
\texttt{duo()} and the corresponding bivariate scales, such as
\texttt{scale\_colour\_bivariate()}. This produces a point-based
bivariate colour map in which glyphs replace polygons as the geographic
display units.

Another glyph design combines a semicircle and a triangle to form a drop
shape \citep{Lucchesi2017}. This can be generated by setting the
\texttt{shape} argument in \texttt{geom\_sf\_glyph()} to
\texttt{"drop"}. Uncertainty is encoded through the rotation angle of
the glyph, ranging from \(-\pi/2\) to \(\pi/2\) for the minimum and
maximum uncertainty values. This mapping is supported by the custom
\texttt{angle} aesthetic and its corresponding
\texttt{scale\_angle\_continuous()} function and guide.

\texttt{ggincerta} also includes glyphs based on Chernoff faces, in
which dimensions of a multivariate observation are mapped to facial
characteristics, such as nose length or mouth curvature
\citep{Chernoff1973}. Although this representation draws on the
familiarity of human facial expressions, its perceptual effectiveness
remains contested \citep{10.1117/12.384865}. When the \texttt{shape}
argument is specified as \texttt{"chernoff"}, glyphs are rendered as
faces using the grob drawing functions provided by the
\texttt{ggChernoff} package. The newly defined \texttt{smile} aesthetic
and its corresponding \texttt{scale\_smile\_continuous()} function allow
uncertainty to be mapped to facial expression.

\subsubsection{Dual map}\label{dual-map}

A dual map combines a glyph map with a colour-filled choropleth map,
implemented through the wrapper layer function
\texttt{geom\_sf\_dualmap()}. The design of this two-layer display
assigns the colour of the central glyph to represent the estimated
value, while the surrounding polygon fill is mapped to uncertainty.
Accordingly, the default fill scale is defined as a greyscale gradient
to avoid competing for visual attention. This design allows the two
variables to be represented through separate mappings, scales, and
guides, such that their original values remain directly interpretable.
When combined with other bivariate scales provided in
\texttt{ggincerta}, the dual map framework is also technically capable
of displaying three or more variables simultaneously, although such
complexity may not constitute effective visualisation practice from a
perceptual perspective.

\begin{figure}

\centering{

\includegraphics[width=1\linewidth,height=\textheight,keepaspectratio]{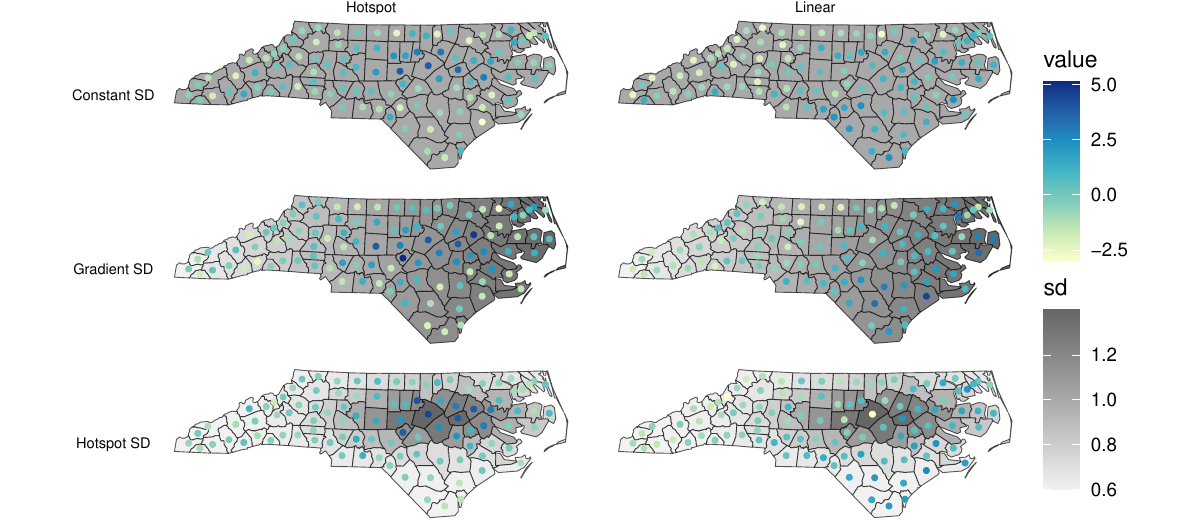}

}

\caption{\label{fig-dualmap-sim}Faceted dual maps of simulated data. The
central glyph colour encodes the simulated value, while the surrounding
fill colour encodes the standard deviation.}

\end{figure}

The drop-shaped glyph map in Figure~\ref{fig-drop-sim} and the dual map
in Figure~\ref{fig-dualmap-sim} both adopt an extrinsic approach to
uncertainty visualisation. Their main difference lies in how visual
attention is distributed across space, as illustrated by the two
simulated examples. In the glyph map, colour continues to carry the
dominant spatial structure of the simulated value. In the dual map,
uncertainty occupies the surrounding areal fill and therefore extends
continuously across each region, giving it a much stronger spatial
presence. As a result, the broader pattern of uncertainty can be
perceived more readily at a glance in the dual map, although this visual
prominence does not necessarily imply greater effectiveness for every
analytical task.

A potential limitation of dual maps is that simultaneous contrast may
affect glyph colour perception, particularly through variations in
background lightness \citep{Brewer01081997}. The default greyscale
background is intended to reduce chromatic interference between the
polygon and glyph colours.

\section{Application}\label{sec-application}

In this section, we demonstrate \texttt{ggincerta} with three real-world
applications: child undernutrition in Bangladesh, fMRI measures from
brain scans, and a factorial experiment for crop improvement project.
Each application highlights particular strengths of \texttt{ggincerta}.

\subsection{Spatial prevalence of child undernutrition in
Bangladesh}\label{spatial-prevalence-of-child-undernutrition-in-bangladesh}

\begin{figure}

\centering{

\includegraphics[width=1\linewidth,height=\textheight,keepaspectratio]{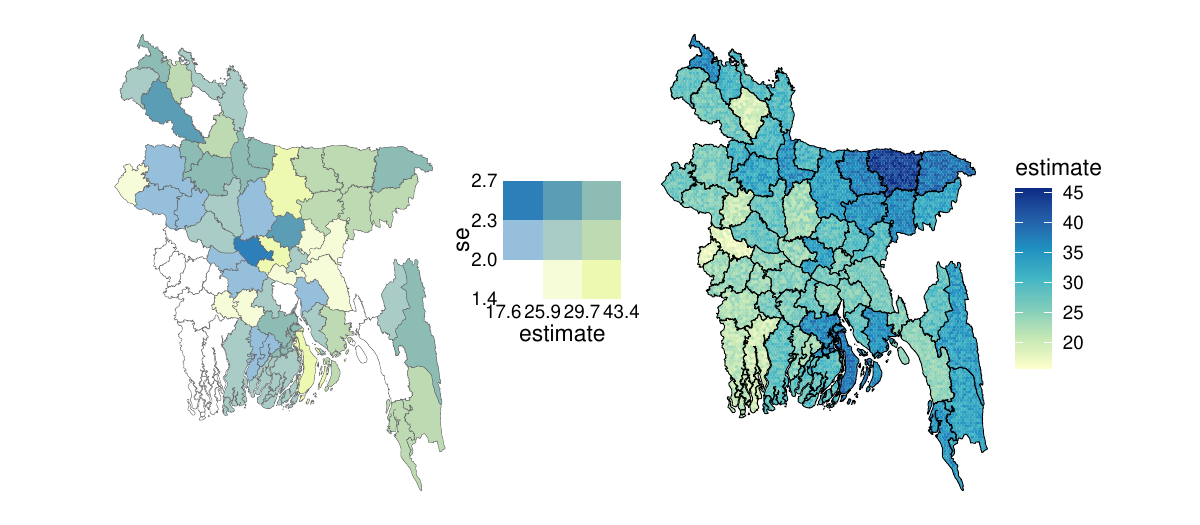}

}

\caption{\label{fig-bangladesh}Bivariate colour map (left) and pixel map
(right) of estimated stunting prevalence and its associated standard
error across districts in Bangladesh.}

\end{figure}

A common form of areal data consists of population health, household
socioeconomic, and other census-related indicators aggregated over
administrative regions. Our first case study considers this common
setting and illustrates how \texttt{ggincerta} can be used to visualise
regional estimates alongside their associated uncertainty. The data are
drawn from a study of child undernutrition in Bangladesh by
\citet{dasPrevalenceChildUndernutrition2022}, which examined several key
indicators, including stunting (low height-for-age), underweight (low
weight-for-age), and wasting (low weight-for-height). They used
model-based small area estimation to obtain more granular,
cross-classified spatio-demographic estimates at the district level,
together with their associated standard errors. To show the spatial
variation in both undernutrition prevalence and estimation uncertainty,
\citet{dasPrevalenceChildUndernutrition2022} presented bivariate
choropleth maps created by \texttt{Vizumap}. Using the supplementary
data, we reproduced the map of stunting with \texttt{ggincerta}, as
shown in Figure~\ref{fig-bangladesh}.

In this case, the bivariate choropleth map provides an overview of
spatial heterogeneity. For example, Figure~\ref{fig-bangladesh} shows
relatively favourable child undernutrition outcomes in southwestern
Bangladesh, where both stunting prevalence and its associated
uncertainty are comparatively low, with higher values becoming more
prominent toward the east and northeast.

However, the bivariate choropleth may overstate spatial variation in
some way. Although approximately 80\% of the standard errors lie between
2.0 and 2.4, small differences may cross bin boundaries and produce
abrupt shifts in the blended colours. The pixel map makes this limited
variation more apparent while retaining the broader spatial pattern in
the estimates. It may therefore provide a more faithful representation
when the objective is to communicate both regional trends and their
associated uncertainty.

\subsection{Trait reliability of task fMRI measures in children across
brain
regions}\label{trait-reliability-of-task-fmri-measures-in-children-across-brain-regions}

Our second application extends spatial visualisation beyond geographical
regions to the cerebral cortex. We use data from a study of the
reliability and longitudinal stability of task-related activation
measured using functional magnetic resonance imaging (fMRI)
\citep{KENNEDY2022119046}. The data comprise 148 cortical parcels
defined by the Destrieux parcellation \citep{DESTRIEUX20101}. For the
n-back 2-back versus 0-back contrast, we visualise the estimated
group-level mean activation and its within-session reliability, derived
from the fixed-effect intercept and variance components of a linear
mixed-effects model, respectively. Although these measures do not form
an estimate--uncertainty pair in the conventional statistical sense,
reliability provides complementary information for interpreting regional
activation.

\begin{figure}

\centering{

\includegraphics[width=1\linewidth,height=\textheight,keepaspectratio]{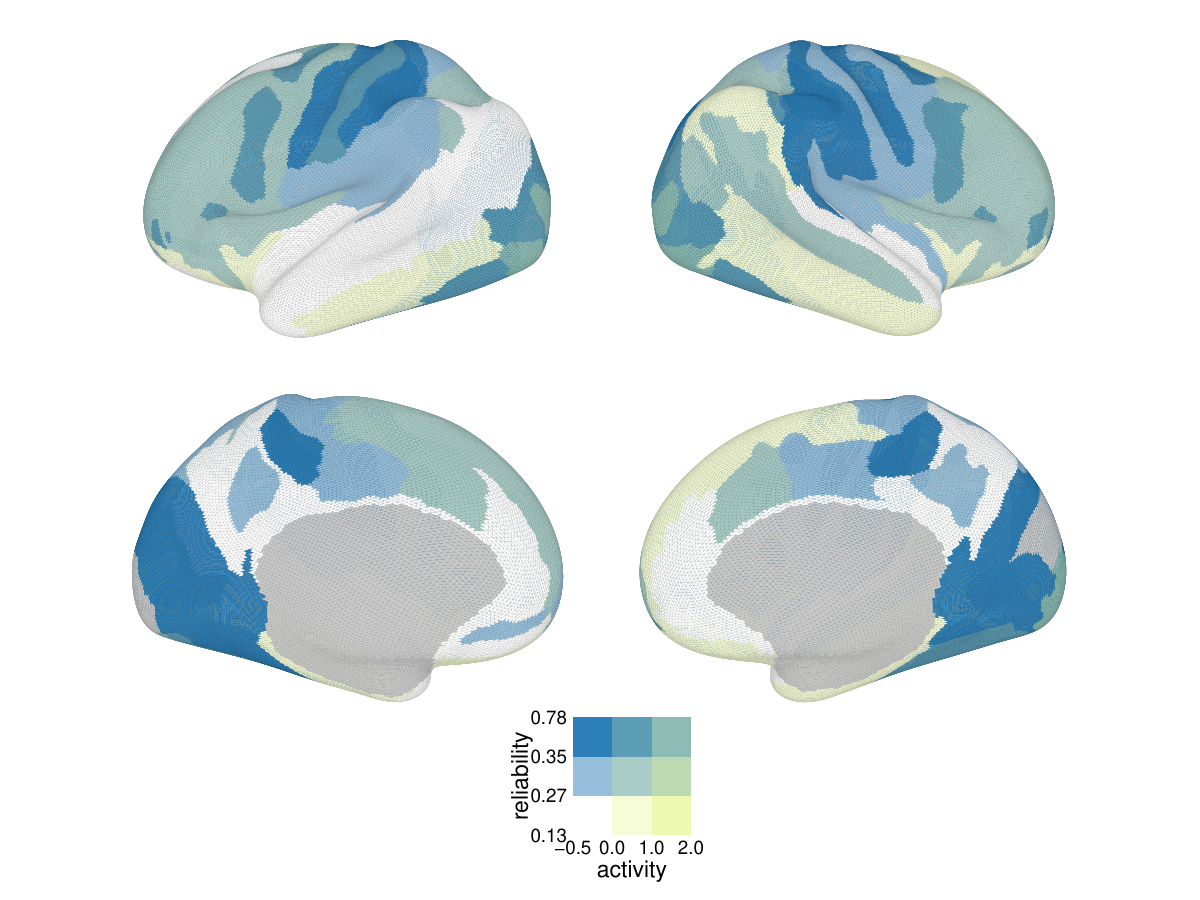}

}

\caption{\label{fig-brain-3d}3D brain plots of regional activity
intensity and reliability mapped using a bivariate colour scale, shown
from medial and lateral views of the left and right hemispheres.}

\end{figure}

\begin{figure}

\centering{

\includegraphics[width=1\linewidth,height=\textheight,keepaspectratio]{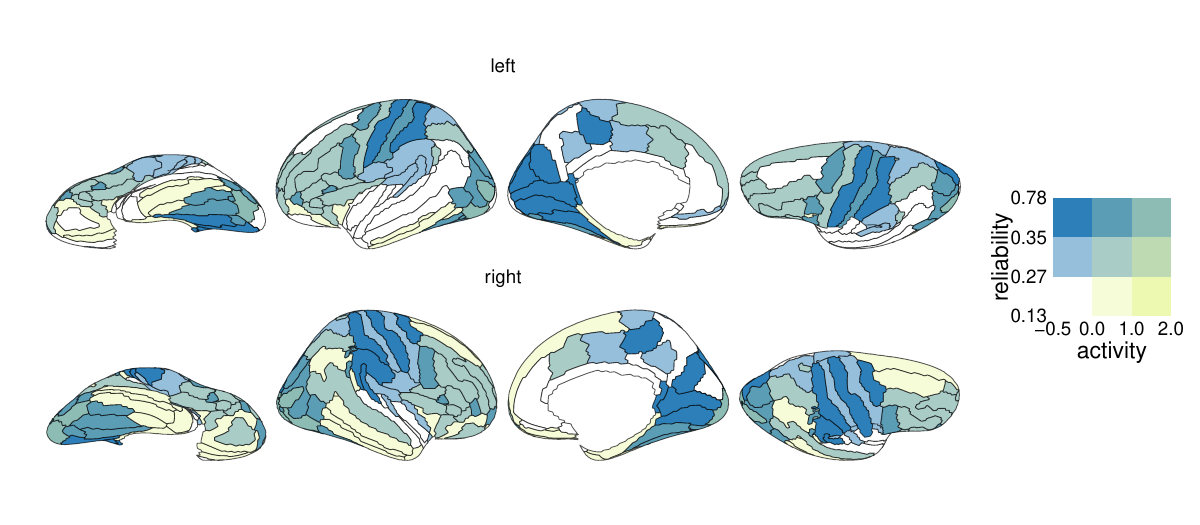}

}

\caption{\label{fig-brain-2d}2D brain plots showing superior, lateral,
medial, and inferior views of the left hemisphere (top row) and right
hemisphere (bottom row).}

\end{figure}

We use \texttt{scale\_fill\_bivariate()} to jointly encode activation
and reliability on two-dimensional cortical maps
(Figure~\ref{fig-brain-2d}) and three-dimensional cortical surfaces
(Figure~\ref{fig-brain-3d}). The 2D atlas is obtained from
\texttt{ggsegFreeSurfer}, while the 3D surfaces are rendered using
\texttt{ggcube}. Both figures use the same tertile breakpoints
calculated from the parcel-level data to ensure consistent colour
mapping. Code \ref{code-3dcortical} illustrates the key steps for
generating the top-left 3D view in Figure~\ref{fig-brain-3d}. The
complete implementation is provided in the Supplementary Materials.

\begin{code}[code-3dcortical]{3D bivariate cortical mapping with ggincerta}

\begin{Shaded}
\begin{Highlighting}[numbers=left,,]
\FunctionTok{ggplot}\NormalTok{(}\FunctionTok{filter}\NormalTok{(brain\_triangles, hemi }\SpecialCharTok{==} \StringTok{"left"}\NormalTok{),}
  \FunctionTok{aes}\NormalTok{(}\AttributeTok{x =}\NormalTok{ x, }\AttributeTok{y =}\NormalTok{ y, }\AttributeTok{z =}\NormalTok{ z,}
      \AttributeTok{group =}\NormalTok{ triangle\_group,}
      \AttributeTok{fill =} \FunctionTok{duo}\NormalTok{(activity, reliability))}
\NormalTok{) }\SpecialCharTok{+}
  \FunctionTok{geom\_polygon\_3d}\NormalTok{() }\SpecialCharTok{+}
  \FunctionTok{scale\_fill\_bivariate}\NormalTok{(}
    \AttributeTok{breaks =} \FunctionTok{list}\NormalTok{(activity\_breaks, reliability\_breaks)}
\NormalTok{  ) }\SpecialCharTok{+}
  \FunctionTok{coord\_3d}\NormalTok{(}\DecValTok{0}\NormalTok{, }\SpecialCharTok{{-}}\DecValTok{90}\NormalTok{, }\SpecialCharTok{{-}}\DecValTok{90}\NormalTok{)}
\end{Highlighting}
\end{Shaded}

\end{code}

Figures \ref{fig-brain-2d} and \ref{fig-brain-3d} reveal distinct
spatial patterns in regional activation and reliability, with
neighbouring cortical parcels often exhibiting similar colour
combinations. Notably, regions with relatively high reliability do not
necessarily exhibit high activation, as reflected by the prominent blue
regions across both hemispheres. Conversely, some regions with higher
activation exhibit relatively low reliability, indicated by the pale
yellow colours. These contrasting combinations demonstrate how bivariate
colour mapping provides additional context for interpreting regional
activation that would not be apparent from an activation map alone.

\subsection{Factorial experimental
designs}\label{factorial-experimental-designs}

The visualisations available in the \texttt{ggincerta} package are not
restricted to spatial data. For instance, the bivariate scale can be
applied to any geometric object. To illustrate its versatility, we
consider a factorial experiment investigating the effects of aluminium
chloride (\(\mathrm{AlCl}_3\)) concentration and acidity, measured by
pH, on canola root growth based on a real consulting problem. The
experiment consists of two treatment factors: pH level and
\(\mathrm{AlCl}_3\) concentration. The treatment is applied to 26 tubs
across two benches.

\begin{figure}[H]

\centering{

\pandocbounded{\includegraphics[keepaspectratio]{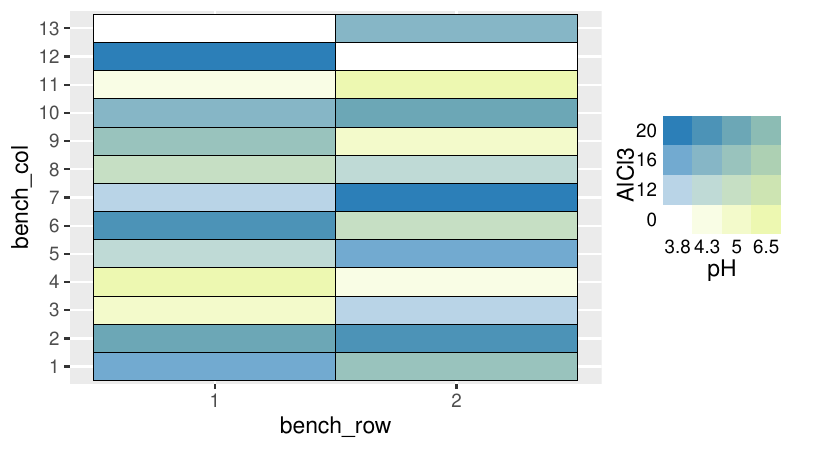}}

}

\caption{\label{fig-exp-design}A factorial experimental design
consisting of two treatment factors: pH level and aluminium chloride
concentration.}

\end{figure}

Figure~\ref{fig-exp-design} applies the bivariate scale to visualise the
application of the two treatment factors. This visualisation helps to
see the distribution of the factorial treatment structure across the
tubs, which can either help to assess the experimental design or provide
a map for conducting the experiment.

\section{Discussion}\label{sec-discuss}

In this paper, we have presented ggincerta, a \texttt{ggplot2} extension
package specialised for spatial uncertainty visualisation. We described
the functions and parameter settings for the supported visualisation
methods and used simulated value--uncertainty patterns to illustrate how
these methods can place different perceptual emphasis on the underlying
information. Compact encodings such as bivariate colour mapping allow
estimates and uncertainty to be represented within a single visual
channel, but can also introduce competition between the two quantities.
Pixel- and glyph-based representations redistribute this visual
emphasis, making different aspects of the estimate or uncertainty more
salient. These differences make the choice of visualisation
task-dependent, motivating multiple options within a common grammar
rather than a single prescribed approach.

A central contribution of \texttt{ggincerta} is its native integration
with the layered grammar of \texttt{ggplot2}. The comparison with
\texttt{Vizumap} illustrates how different levels of integration can
affect functionality and usability. Rather than prescribing a fixed set
of graphical components, \texttt{ggincerta} defines only those required
by each visual encoding and inherits the remaining functionality from
\texttt{ggplot2}. This approach ensures consistency with the existing
system while facilitating the reuse and composition of components across
the broader \texttt{ggplot2} extension ecosystem.

While our simulated examples and case studies demonstrate the
flexibility of the supported methods, they do not establish which
methods or parameter settings best support particular analytical tasks.
Recent experimental work by \citet{mason2026colourblindednoise} examines
uncertainty visualisation through signal suppression, highlighting the
importance of evaluating how visual encodings affect the perception of
statistical signals. An important avenue for future research is
therefore to systematically investigate how data patterns, visualisation
methods, and design choices interact to influence visual interpretation.
The consistent framework provided by \texttt{ggincerta} facilitates such
comparisons, which could inform practical design guidance and future
refinements of the visualisation methods.

\section*{Acknowledgments}\label{acknowledgments}
\addcontentsline{toc}{section}{Acknowledgments}

Weihao Li acknowledges support from the Grains Research and Development
Corporation (GRDC) as part of the Analytics for the Australian Grains
Industry (AAGI) initiative (UOA2301-005OPX).

\section{Supplementary Material}\label{supplementary-material}

\begin{description}
\item[Figure reproduction code:]
Complete R code for reproducing all figures presented in the article,
including simulated examples and three applications. (PDF file)
\item[Supporting datasets:]
All datasets used to generate the figures in the article, including
simulated spatial data and datasets from the three applications. (ZIP
file)
\end{description}

\section*{Computational details}\label{computational-details}
\addcontentsline{toc}{section}{Computational details}

The R package \texttt{ggincerta} \citep{ggincerta} is available on
\href{https://cran.r-project.org/package=ggincerta}{CRAN}, with its
source code hosted on
\href{https://github.com/maggiexma/ggincerta}{GitHub}. All data and R
code required to reproduce the figures are provided in the supplementary
material, and the article was prepared using Quarto.

We used R v. 4.6.0 \citep{base} and the following R packages: ggcube v.
0.1.0 \citep{ggcube}, ggpattern v. 1.3.1 \citep{ggpattern}, patchwork v.
1.3.2 \citep{patchwork}, sf v. 1.1.3 \citep{sf2018, sf2023}, tidyverse
v. 2.0.0 \citep{tidyverse}.

\bibliography{references.bib,grateful-refs.bib}

@Manual{base,
title = {{R}: A Language and Environment for Statistical Computing},
  author = {{R Core Team}},
  organization = {R Foundation for Statistical Computing},
  address = {Vienna, Austria},
  year = {2026},
  doi = {10.32614/R.manuals},
  url = {https://www.R-project.org/},
}

@Manual{ggcube,
title = {{ggcube}: 3D Plotting with `{ggplot2}'},
  author = {Matthew Kling},
  year = {2026},
  note = {R package version 0.1.0},
  url = {https://CRAN.R-project.org/package=ggcube},
  doi = {10.32614/CRAN.package.ggcube},
}

@Manual{ggpattern,
title = {{ggpattern}: `{ggplot2}' Pattern Geoms},
  author = {Mike FC and Trevor L. Davis and {ggplot2 authors}},
  year = {2026},
  note = {R package version 1.3.1},
  url = {https://CRAN.R-project.org/package=ggpattern},
  doi = {10.32614/CRAN.package.ggpattern},
}

@Manual{patchwork,
title = {{patchwork}: The Composer of Plots},
  author = {Thomas Lin Pedersen},
  year = {2025},
  note = {R package version 1.3.2},
  url = {https://CRAN.R-project.org/package=patchwork},
  doi = {10.32614/CRAN.package.patchwork},
}

@Book{sf2023,
  author = {Edzer Pebesma and Roger Bivand},
  title = {{Spatial Data Science: With applications in R}},
  year = {2023},
  publisher = {{Chapman and Hall/CRC}},
  url = {https://r-spatial.org/book/},
  doi = {10.1201/9780429459016},
}

@Article{sf2018,
  author = {Edzer Pebesma},
  title = {{Simple Features for R: Standardized Support for Spatial Vector Data}},
  year = {2018},
  journal = {{The R Journal}},
  doi = {10.32614/RJ-2018-009},
  url = {https://doi.org/10.32614/RJ-2018-009},
  pages = {439--446},
  volume = {10},
  number = {1},
}

@Article{tidyverse,
  title = {Welcome to the {tidyverse}},
  author = {Hadley Wickham and Mara Averick and Jennifer Bryan and Winston Chang and Lucy D'Agostino McGowan and Romain François and Garrett Grolemund and Alex Hayes and Lionel Henry and Jim Hester and Max Kuhn and Thomas Lin Pedersen and Evan Miller and Stephan Milton Bache and Kirill Müller and Jeroen Ooms and David Robinson and Dana Paige Seidel and Vitalie Spinu and Kohske Takahashi and Davis Vaughan and Claus Wilke and Kara Woo and Hiroaki Yutani},
  year = {2019},
  journal = {Journal of Open Source Software},
  volume = {4},
  number = {43},
  pages = {1686},
  doi = {10.21105/joss.01686},
}

@article{Lucchesi2021, doi = {10.21105/joss.02409}, url = {https://doi.org/10.21105/joss.02409}, year = {2021}, publisher = {The Open Journal}, volume = {6}, number = {59}, pages = {2409}, author = {Lucchesi, Lydia R. and Kuhnert, Petra M. and Wikle, Christopher K.}, title = {Vizumap: an {R} package for visualising uncertainty in spatial data}, journal = {Journal of Open Source Software} }

@book{leland_wilkinson_grammar_2005,
	address = {New York},
	series = {Statistics and {Computing}},
	title = {The {Grammar} of {Graphics}},
	copyright = {http://www.springer.com/tdm},
	isbn = {978-0-387-24544-7},
	url = {http://link.springer.com/10.1007/0-387-28695-0},
	language = {en},
	urldate = {2025-08-20},
	publisher = {Springer-Verlag},
	author = {Wilkinson, Leland},
	year = {2005},
	doi = {10.1007/0-387-28695-0},
}

@article{MacEachren01012005,
author = {Alan M. MacEachren and Anthony Robinson and Susan Hopper and Steven Gardner and Robert Murray and Mark Gahegan and Elisabeth Hetzler},
title = {Visualizing Geospatial Information Uncertainty: What We Know and What We Need to Know},
journal = {Cartography and Geographic Information Science},
volume = {32},
number = {3},
pages = {139--160},
year = {2005},
publisher = {Taylor \& Francis},
doi = {10.1559/1523040054738936},
}

@article{Lucchesi2017,
author = {Lucchesi, Lydia R. and Wikle, Christopher K.},
title = {Visualizing uncertainty in areal data with bivariate choropleth maps, map pixelation and glyph rotation},
journal = {Stat},
volume = {6},
number = {1},
pages = {292-302},
doi = {10.1002/sta4.150},
url = {https://onlinelibrary.wiley.com/doi/abs/10.1002/sta4.150},
eprint = {https://onlinelibrary.wiley.com/doi/pdf/10.1002/sta4.150},
year = {2017}
}

@misc{prener_biscale_2022,
	title = {biscale: {Tools} and {Palettes} for {Bivariate} {Thematic} {Mapping}},
	copyright = {GPL-3},
	shorttitle = {biscale},
	url = {https://cran.r-project.org/web/packages/biscale/index.html},
	urldate = {2025-07-30},
	author = {Prener, Christopher and Grossenbacher, Timo and Zehr, Angelo and Stevens, Joshua},
	month = may,
	year = {2022},
}

@misc{taylor_pixelate_2020,
	title = {Pixelate to communicate: visualising uncertainty in maps of disease risk and other spatial continua},
	shorttitle = {Pixelate to communicate},
	url = {http://arxiv.org/abs/2005.11993},
	doi = {10.48550/arXiv.2005.11993},
	urldate = {2025-10-07},
	publisher = {arXiv},
	author = {Taylor, Aimee R. and Watson, James A. and Buckee, Caroline O.},
	month = may,
	year = {2020},
	note = {arXiv:2005.11993 [stat]},
}

@inproceedings{correll_value-suppressing_2018,
	address = {New York, NY, USA},
	series = {{CHI} '18},
	title = {Value-{Suppressing} {Uncertainty} {Palettes}},
	isbn = {978-1-4503-5620-6},
	url = {https://doi.org/10.1145/3173574.3174216},
	doi = {10.1145/3173574.3174216},
	urldate = {2025-09-24},
	booktitle = {Proceedings of the 2018 {CHI} {Conference} on {Human} {Factors} in {Computing} {Systems}},
	publisher = {Association for Computing Machinery},
	author = {Correll, Michael and Moritz, Dominik and Heer, Jeffrey},
	year = {2018},
	pages = {1--11},
}

@book{Bertin2010SemiologyOG,
  author = {Bertin, Jacques},
  title = {Semiology of Graphics: Diagrams, Networks, Maps},
  translator = {Berg, William J.},
  year = {2010},
  publisher = {Esri Press},
  isbn = {9781589482616},
  url = {https://www.esri.com/en-us/esri-press/browse/semiology-of-graphics-diagrams-networks-maps}
}

@ARTICLE{6327283,
  author={Boukhelifa, Nadia and Bezerianos, Anastasia and Isenberg, Tobias and Fekete, Jean-Daniel},
  journal={IEEE Transactions on Visualization and Computer Graphics},
  title={Evaluating Sketchiness as a Visual Variable for the Depiction of Qualitative Uncertainty},
  year={2012},
  volume={18},
  number={12},
  pages={2769-2778},
  doi={10.1109/TVCG.2012.220}
}

@book{murrell-hdvw-2026-03-24,
    title = {How Data Visualisation Works},
    author = {Murrell, Paul},
    note = {Version 0.9 (build 2026-03-24)},
    year = 2026,
    publisher = {The University of Auckland}
}

@book{stone_field_2016,
	address = {New York},
	title = {A {Field} {Guide} to {Digital} {Color}},
	isbn = {978-0-429-18740-7},
	doi = {10.1201/b12887},
	publisher = {A K Peters/CRC Press},
	author = {Stone, Maureen},
	month = apr,
	year = {2016},
}

@incollection{Brewer1994,
	series = {Modern {Cartography} {Series}},
	title = {Color Use Guidelines for Mapping and Visualization},
	volume = {2},
	issn = {1363-0814},
	url = {https://www.sciencedirect.com/science/article/pii/B9780080424156500144},
	doi = {10.1016/B978-0-08-042415-6.50014-4},
	booktitle = {Visualization in {Modern} {Cartography}},
	publisher = {Academic Press},
	author = {Brewer, Cynthia A.},
	editor = {MacEachren, Alan M. and Taylor, D. R. Fraser},
	year = {1994},
	pages = {123--147},
}

@article{evansDynamicDisplaySpatial1997a,
  title = {Dynamic Display of Spatial Data-Reliability: {{Does}} It Benefit the Map User?},
  author = {Evans, Beverley J.},
  year = 1997,
  journal = {Computers \& Geosciences},
  volume = {23},
  number = {4},
  pages = {409--422},
  issn = {0098-3004},
  doi = {10.1016/S0098-3004(97)00011-3}
}

@article{doi:10.3138/B204-32P4-263L-76W0,
  title = {Visualizing Uncertainty in Soil Maps by Animation},
  author = {Fisher, Peter F},
  year = 1993,
  journal = {Cartographica},
  volume = {30},
  number = {2-3},
  eprint = {https://doi.org/10.3138/B204-32P4-263L-76W0},
  pages = {20--27},
  doi = {10.3138/B204-32P4-263L-76W0}
}

@article{Carr01011992,
  title = {Hexagon Mosaic Maps for Display of Univariate and Bivariate Geographical Data},
  author = {Carr, Daniel B. and Olsen, Anthony R. and White, Denis},
  year = 1992,
  journal = {Cartography and Geographic Information Systems},
  volume = {19},
  number = {4},
  eprint = {https://doi.org/10.1559/152304092783721231},
  pages = {228--236},
  publisher = {Taylor \& Francis},
  doi = {10.1559/152304092783721231}
}

@article{Sahr01012003,
  title = {Geodesic Discrete Global Grid Systems},
  author = {Sahr, Kevin and White, Denis and Kimerling, A. Jon},
  year = 2003,
  journal = {Cartography and Geographic Information Science},
  volume = {30},
  number = {2},
  eprint = {https://doi.org/10.1559/152304003100011090},
  pages = {121--134},
  publisher = {Taylor \& Francis},
  doi = {10.1559/152304003100011090}
}

@article{Pang1997,
  title = {Approaches to Uncertainty Visualization},
  author = {Pang, Alex and Wittenbrink, Craig and Lodha, Suresh},
  year = 1997,
  month = nov,
  journal = {The Visual Computer},
  volume = {13},
  number = {8},
  pages = {370--390},
  doi = {10.1007/s003710050111}
}

@article{Chernoff1973,
 ISSN = {01621459, 1537274X},
 URL = {http://www.jstor.org/stable/2284077},
 author = {Herman Chernoff},
 journal = {Journal of the American Statistical Association},
 number = {342},
 pages = {361--368},
 publisher = {[American Statistical Association, Taylor & Francis, Ltd.]},
 title = {The Use of Faces to Represent Points in K-Dimensional Space Graphically},
 urldate = {2026-05-25},
 volume = {68},
 year = {1973}
}

@Manual{Mason2026,
    title = {ggdibbler: Add Uncertainty to Data Visualisations},
    author = {Harriet Mason and Dianne Cook and Sarah Goodwin and Susan VanderPlas},
    year = {2026},
    note = {R package version 0.6.5},
    url = {https://CRAN.R-project.org/package=ggdibbler},
    doi = {10.32614/CRAN.package.ggdibbler},
}

@Book{Wickham2016,
    author = {Hadley Wickham},
    title = {ggplot2: Elegant Graphics for Data Analysis},
    publisher = {Springer-Verlag New York},
    year = {2016},
    isbn = {978-3-319-24277-4},
    url = {https://ggplot2.tidyverse.org},
}

@book{noelaccressieStatisticsSpatialData2015,
  author = {Cressie, Noel},
  title = {Statistics for Spatial Data, Revised Edition},
  year = {2015},
  edition = {2},
  series = {Wiley Classics Library},
  publisher = {John Wiley \& Sons},
  isbn = {9781119114611}
}

@article{dasPrevalenceChildUndernutrition2022,
  title = {Prevalence of Child Undernutrition Measures and Their Spatio-Demographic Inequalities in {{Bangladesh}}: An Application of Multilevel {{Bayesian}} Modelling},
  author = {Das, Sumonkanti and Baffour, Bernard and Richardson, Alice},
  year = 2022,
  month = may,
  journal = {BMC Public Health},
  volume = {22},
  number = {1},
  pages = {1008},
  issn = {1471-2458},
  doi = {10.1186/s12889-022-13170-4}
}

@article{KENNEDY2022119046,
title = {Reliability and stability challenges in {ABCD} task {fMRI} data},
journal = {NeuroImage},
volume = {252},
pages = {119046},
year = {2022},
issn = {1053-8119},
doi = {10.1016/j.neuroimage.2022.119046},
url = {https://www.sciencedirect.com/science/article/pii/S1053811922001756},
author = {James T. Kennedy and Michael P. Harms and Ozlem Korucuoglu and Serguei V. Astafiev and Deanna M. Barch and Wesley K. Thompson and James M. Bjork and Andrey P. Anokhin}
}

@article{DESTRIEUX20101,
title = {Automatic parcellation of human cortical gyri and sulci using standard anatomical nomenclature},
journal = {NeuroImage},
volume = {53},
number = {1},
pages = {1-15},
year = {2010},
issn = {1053-8119},
doi = {10.1016/j.neuroimage.2010.06.010},
url = {https://www.sciencedirect.com/science/article/pii/S1053811910008542},
author = {Christophe Destrieux and Bruce Fischl and Anders Dale and Eric Halgren}
}

@book{banerjee2014hierarchical,
  title     = {Hierarchical Modeling and Analysis for Spatial Data},
  author    = {Banerjee, Sudipto and Carlin, Bradley P. and Gelfand, Alan E.},
  series    = {Chapman \& Hall/CRC Monographs on Statistics and Applied Probability},
  edition   = {2},
  year      = {2014},
  publisher = {CRC Press},
  isbn      = {9781439819180}
}

@incollection{LI2018313,
  title = {1.22 - Spatial Data Uncertainty},
  booktitle = {Comprehensive Geographic Information Systems},
  author = {Li, Linna and Ban, Hyowon and Wechsler, Suzanne P. and Xu, Bo},
  editor = {Huang, Bo},
  year = 2018,
  pages = {313--340},
  publisher = {Elsevier},
  address = {Oxford},
  doi = {10.1016/B978-0-12-409548-9.09610-X},
  isbn = {978-0-12-804793-4}
}

@article{MacEachren_1992,
  title = {Visualizing Uncertain Information},
  author = {MacEachren, Alan M.},
  year = 1992,
  month = jun,
  journal = {Cartographic Perspectives},
  number = {13},
  pages = {10--19},
  doi = {10.14714/CP13.1000}
}

@ARTICLE{5963660,
  author={Healey, Christopher and Enns, James},
  journal={IEEE Transactions on Visualization and Computer Graphics},
  title={Attention and Visual Memory in Visualization and Computer Graphics},
  year={2012},
  volume={18},
  number={7},
  pages={1170-1188},
  doi={10.1109/TVCG.2011.127}
}

@ARTICLE{6327245,
  author={Haroz, Steve and Whitney, David},
  journal={IEEE Transactions on Visualization and Computer Graphics},
  title={How Capacity Limits of Attention Influence Information Visualization Effectiveness},
  year={2012},
  volume={18},
  number={12},
  pages={2402-2410},
  doi={10.1109/TVCG.2012.233}
}

@ARTICLE{6327255,
  author={MacEachren, Alan M. and Roth, Robert E. and O'Brien, James and Li, Bonan and Swingley, Derek and Gahegan, Mark},
  journal={IEEE Transactions on Visualization and Computer Graphics},
  title={Visual semiotics \& Uncertainty Visualization: An Empirical Study},
  year={2012},
  volume={18},
  number={12},
  pages={2496-2505},
  doi={10.1109/TVCG.2012.279}
}

@article{Kinkeldey01112014,
author = {Christoph Kinkeldey and Alan M. MacEachren and Jochen Schiewe},
title = {How to Assess Visual Communication of Uncertainty? A Systematic Review of Geospatial Uncertainty Visualisation User Studies},
journal = {The Cartographic Journal},
volume = {51},
number = {4},
pages = {372--386},
year = {2014},
publisher = {Taylor \& Francis},
doi = {10.1179/1743277414Y.0000000099},
URL = {https://doi.org/10.1179/1743277414Y.0000000099},
eprint = {https://doi.org/10.1179/1743277414Y.0000000099}
}

@book{munznerVisualizationAnalysisDesign2014,
  title = {Visualization {{Analysis}} and {{Design}}},
  author = {Munzner, Tamara},
  year = 2014,
  month = dec,
  publisher = {A K Peters/CRC Press},
  address = {New York},
  doi = {10.1201/b17511},
  isbn = {978-0-429-08890-2}
}

@article{kinkeldeyEvaluatingEffectVisually2017,
  title = {Evaluating the Effect of Visually Represented Geodata Uncertainty on Decision-Making: Systematic Review, Lessons Learned, and Recommendations},
  shorttitle = {Evaluating the Effect of Visually Represented Geodata Uncertainty on Decision-Making},
  author = {Kinkeldey, Christoph and MacEachren, Alan M. and Riveiro, Maria and Schiewe, Jochen},
  year = 2017,
  month = jan,
  journal = {Cartography and Geographic Information Science},
  volume = {44},
  number = {1},
  pages = {1--21},
  publisher = {Taylor \& Francis},
  issn = {1523-0406},
  doi = {10.1080/15230406.2015.1089792},
  urldate = {2026-10-04}
}

@inproceedings{10.1117/12.384865,
author = {Christopher J. Morris and David S. Ebert and Penny L. Rheingans},
title = {{Experimental analysis of the effectiveness of features in Chernoff faces}},
volume = {3905},
booktitle = {28th AIPR Workshop: 3D Visualization for Data Exploration and Decision Making},
editor = {William R. Oliver},
organization = {International Society for Optics and Photonics},
publisher = {SPIE},
pages = {12 -- 17},
year = {2000},
doi = {10.1117/12.384865},
URL = {https://doi.org/10.1117/12.384865}
}

@misc{mason2026colourblindednoise,
      title={Colour Blinded by the Noise},
      author={Harriet Mason and Rachel Rogers and Alison Kleffner and Dianne Cook},
      year={2026},
      eprint={2608.17976},
      archivePrefix={arXiv},
      primaryClass={cs.HC},
      url={https://arxiv.org/abs/2608.17976},
}

@Manual{ggincerta,
    author = {Xueqi Ma and Emi Tanaka},
    title = {{ggincerta}: Extend `{ggplot2}' with Layers and Scales for Spatial Uncertainty Visualization},
    year = {2026},
    note = {R package version 0.2.1},
    url = {https://CRAN.R-project.org/package=ggincerta},
}

@Article{Wickham2010,
    author = {Hadley Wickham},
    title = {A Layered Grammar of Graphics},
    journal = {Journal of Computational and Graphical Statistics},
    year = {2010},
    volume = {19},
    number = {1},
    pages = {3--28},
    doi = {10.1198/jcgs.2009.07098},
}

@article{Brewer01081997,
author = {Brewer, Cynthia A.},
title = {Evaluation of a Model for Predicting Simultaneous Contrast on Color Maps},
journal = {The Professional Geographer},
volume = {49},
number = {3},
pages = {280--294},
year = {1997},
publisher = {Routledge},
doi = {10.1111/0033-0124.00077},
URL = {https://doi.org/10.1111/0033-0124.00077},
eprint = {https://doi.org/10.1111/0033-0124.00077}
}

\end{document}